\documentstyle[11pt,epsf]{article}
\newcommand{\sptwo}{1.6}

\newcommand{\doublespace}{\edef\baselinestretch{\sptwo}\Large\normalsize}

\begin{document}

\doublespace

\begin{center}
{\large \bf {\it Welcher Weg} Experiments, Duality and  the Orthodox
Bohr's Complementarity Principle}
\end{center}

\medskip

\begin{center}
{\bf S. Bandyopadhyay$\footnote{Corresponding author. E-mail: bandy@quantum1.unl.edu}$}\\
{\it Department of Electrical Engineering,
University of Nebraska,
Lincoln, Nebraska 68588-0511, USA}
\end{center}

\medskip

\begin{abstract}
In its  most orthodox form,    Bohr's Complementarity Principle 
states that a quanton (a quantum system consisting of a Boson or Fermion)
can either behave as a particle or as wave, but never simultaneosuly as both.
A less orthodox interpretation of this Principle is the ``duality condition'' embodied in a 
mathematical inequality due to Englert [B-G Englert,  Phys. Rev. Lett., \underline{ 77}, 2154 (1996)]
which allows wave and particle attributes to co-exist, but postulates
that a stronger manifestation of the particle nature leads to a 
weaker manifestation of the wave nature and vice versa.
In this Letter, we show that some recent {\it welcher weg} ("which 
path") experiments in interferometers and similar set-ups, that
claim to have validated, or invalidated, the Complementarity 
Principle, actually shed no light on the orthodox interpretation.
They may have instead validated the weaker duality condition, but even that is not completely obvious.
We propose simple modifications
to these experiments which we believe can test 
the orthodox Complementarity Principle and also shed light on the nature of 
wavefunction collapse and quantum erasure.

\end{abstract}


\twocolumn

The orthodox Bohr's complementarity principle \cite{bohr} states that  a quanton can behave 
 {\it either} as a particle {\it or} as a wave, but never as {\it  both at the same time}.
{\it Welcher weg} experiments conducted with two-path interferometers 
(or analogous set-ups) are a suitable vehicle to test
this strict complementarity between the wave- and particle-nature. If one can determine - 
even in principle - which of the two paths in the interferometer was traversed by 
the quanton, then the entity behaves as a ``particle'' since a ``wave'' would have 
sampled both paths simultaneously. In this case, there should be no
interference (or any other wave-like behavior) if the orthodox Complementarity 
Principle holds. On the other hand, if there is interference, then the 
wave property is intact in which case it should have been impossible to 
discern the particle attribute, i.e. to tell which path was traversed. 
The orthodox Complementarity principle therefore allows only sharp wave 
or sharp particle attribute, but not both.
A somewhat tempered version of the Complementarity Principle is the
{\it duality principle} due to Englert \cite{englert} which 
states that a quantum system can simultaneosuly exhibit  wave and particle 
behavior, but sharperning of the wave character blurs the particle character and vice versa. In fact,
Englert derives an inequality 
\begin{equation}
P^2 + V^2 \leq 1
\end{equation}
where $P$ is a measure of the ``which-path'' information (particle 
attribute) and $V$ is a 
measure of the ``(interference) fringe-visibility'' (wave attribute). Equation (1)
immediately shows that stronger wave or stronger particle behavior can be manifested only 
at the expense of each other.

\medskip

\noindent {\bf Welcher Weg experiments that question the orthodox Complementarity 
Principle}

\medskip

Experiments purported to demonstrate violation of the 
{\it orthodox} version of the Complementarity
Principle were proposed and carried out in the past.
Ghose, Home and Agarwal \cite{ghose} had proposed a biprism
experiment schematically depicted in Fig. 1(a). A single photon 
source emits a single photon which is split into orthogonal 
states $\psi_r$ and $\psi_t$ by a 50:50 beam splitter. They
are detected by two photon detectors $D_r$ and $D_t$. If the 
photon behaves truly as a particle, then it should 
 be detected at either $D_r$ or $D_t$ (but never at both)
since a particle cannot traverse two paths simultaneously.
That is, there should be perfect
anti-coincidence between $D_r$ and $D_t$, or, in other words, {\it
either} $D_r$ {\it or} $D_r$ will click but {\it both}
will never click 
in between the arrival of two successive
photons. The clever twist in this experiment, motivated 
by an experiment performed in the 19th century by Jagdish Chandra
Bose \cite{bose}, 
is the placement of the biprism with a 
small tunneling gap in the path of the transmitted photon.
If $D_t$ clicks {\it and $D_r$ does not}, then we have made
a ``which path'' determination (the particle took the
path of transmission as opposed to reflection) and a {\it sharp} particle nature
is demonstrated \cite{aspect}.
Yet, to arrive at $D_t$, the particle must have tunneled 
through the biprism and tunneling is a {\it sharp} wave attribute. In this 
experiment, later conducted by Mizobuchi and Ohtake \cite{mizobuchi},
perfect anti-coincidence was found between $D_t$ and $D_r$
demonstrating the particle nature. Yet, the very fact that $D_t$
ever clicked required tunneling and hence the existence of a 
wave nature. It was claimed  that in this experiment, a 
photon was behaving both sharply as a particle and as a wave 
in violation of Bohr's 
Complementarity Principle. A slight modification of this 
experiment has been proposed by Rangwala and Roy \cite{rangwala}
where interference is used instead of tunneling to showcase
the wave-like behavior. They claimed that quantum mechanics 
does not prohibit the demonstration of simultaneous
wave and particle behavior; rather, it prohibits their 
simultaneity only when the wave and particle attributes are ``complementary'' \cite{dirac} in the sense that projection operators associated with them
do not {\it commute}. This is actually consistent with Englert's
work \cite{englert} in that Englert takes pain to point out
that Equation (1) does not rely on Heisneberg type uncertainty, i.e.
the following relation need {\it not} hold:
\begin{equation}
\Delta P \Delta V \geq {{1}\over{2}} |<[P,V]>| ~.
\end{equation}

In the experiments
of refs. \cite{ghose, mizobuchi,
rangwala}, the wave and particle behavior  supposedly are not truly complementary and hence not subject to the restrictions of the 
orthodox Complementarity Principle.

\begin{figure*}
\epsfxsize=5.8in
\epsfysize=6.3in
\centerline{\epsffile{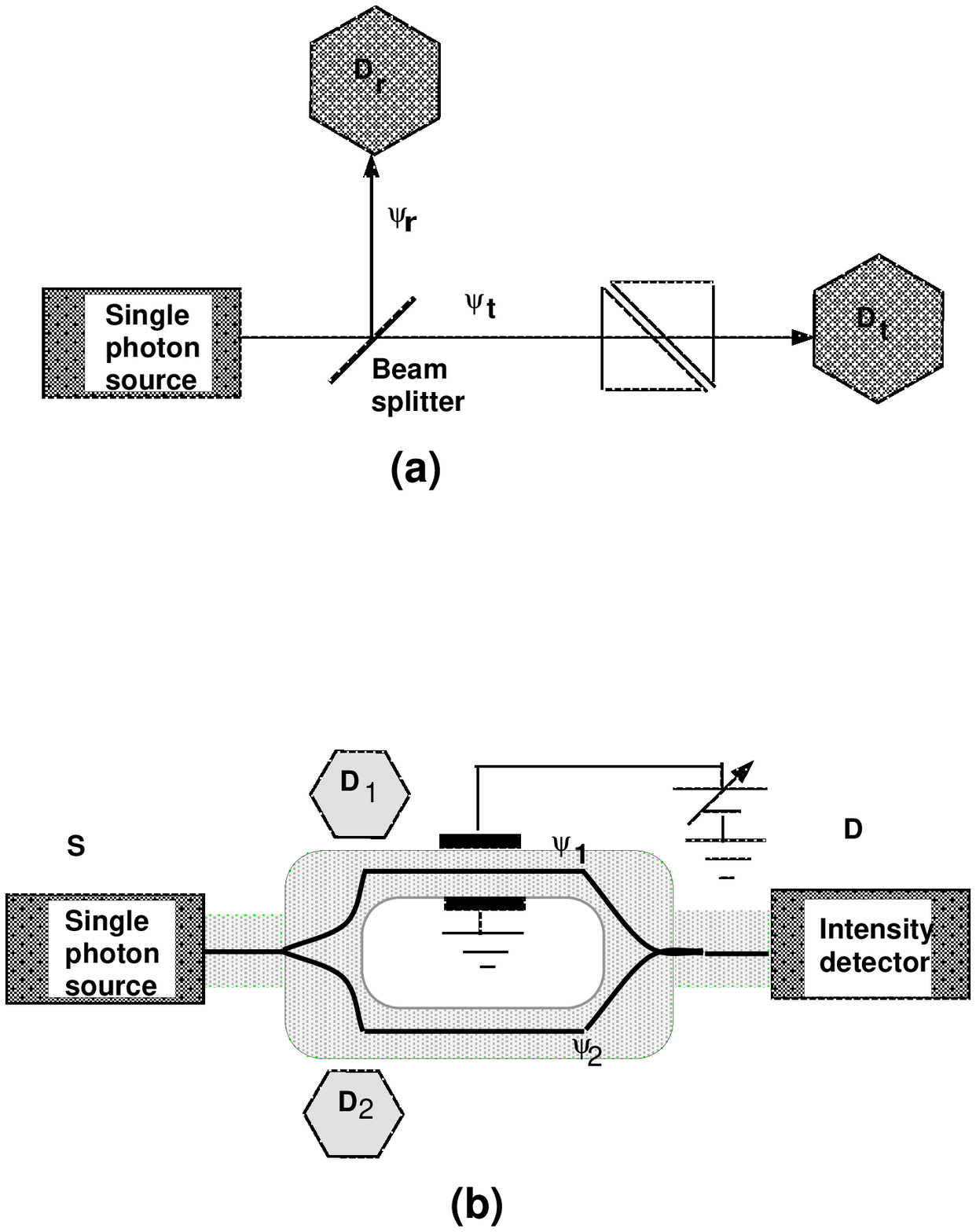}}
\caption{(a) A welcher-weg biprism experiment to demonstrate violation of the 
Complementarity Principle; (b) a welcher-weg Mach-Zender interferometry experiment
that can test the Complementarity Principle more rigorously.}
\end{figure*}

A mathematical framework to determine true ``complementarity'' 
between wave and particle attributes was first addressed by Kar. et. al. \cite{kar}.
Complementary observables as those whose projection
operators {\it do not commute}, i.e. have no common 
eigenvectors.  In the experiments of refs. \cite{ghose, mizobuchi,
rangwala}, the Hilbert spaces $H_r$ and $H_t$ associated with the reflected state 
$\psi_r$ and the transmitted state $\psi_t$ are orthogonal (since there is always anti-coincidence 
between reflection and transmission). Hence the
projection operators $P_r$ and $P_t$, corresponding to 
reflection and transmission respectively,  always commute. If we assume that the
wave property (tunneling) is represented by some projection
operator $P_{wave}$, its Hilbert space is contained within the 
Hilbert space of $P_t$. i.e. $<\psi|P_{wave}|\psi>$ $\leq$  $<\psi|P_{t}|\psi>$
and is hence orthogonal to $H_r$. Thus, $P_{wave}$ commutes with $P_r$. But $P_{wave}$
must also commute with $P_t$ since every state $\psi_t$ that tunnels
through the biprism and reaches $D_t$ is a common eigenvector of 
these two operators. Thus, $P_{wave}$, $P_t$ and $P_r$ all commute
with each other. Hence the sharp wave property (tunneling) and the 
sharp particle
property (anti-coincidence between detectors $D_t$ and $D_r$) are 
{\it not} complementary and their simultaneous observation is
{\it not} prohibited by the Complementarity Principle (equation (1) however, 
must still be obeyed, but the experiments did not test this inequality).
In concluding their paper, Kar et. al. \cite{kar} point out that 
the experiments of refs. \cite{ghose, mizobuchi, rangwala} do
not test wave and particle properties that are complementary
and hence can draw no conclusion about the validity or invalidity of the ``orthodox'' Complementarity Principle.

\medskip

\noindent {\bf An alternate {\it welcher weg} experiment to test the orthodox Complementarity Principle}

\medskip

We propose an alternate {\it welcher weg} 
experiment where sharp particle behavior and sharp 
wave behavior would be complementary in the sense defined 
by Kar and co-workers. Consequently, this is an {\it unambiguous}
experiment where simultaneous exhibition of sharp particle- 
and wave-character will give lie to the orthodox version of the Complementarity Principle.
We point out that these experiments are worth conducting since their
outcome is by no means a foregone conclusion. The orthodox Complementarity 
Principle is not sacrosanct (even though it is viewed by some as a cornerstone
of the Copenhagen interpretation of quantum mechanics); viewpoints 
due to Einstein and DeBroglie do not subscribe to the Complementarity
Principle \cite{einstein, debroglie}. 

Consider a Mach-Zender type interferometer
as shown in Fig. 1(b). 
Proximity photon detectors $D_1$ and $D_2$ are placed near each limb as shown.
A single photon source injects photons at terminal $S$ which reaches 
a screen $D$ after traveling along the two possible paths comprising the 
arms of the ring (actually there are a denumerably infinite 
number of paths possible if we take into account multiple 
reflections, but they are not important in this context).

To demonstrate an invalidation of the orthodox Complementarity Principle,
we need to demonstrate {\it two} effects simultaneously:

\begin{enumerate}

\item Perfect anti-coincidence between $D_1$ and $D_2$ (sharp particle
nature).

\item Existence of an interference pattern at $D$ (sharp wave nature)

\end{enumerate}

In this example, the projection operators $P_1$ and $P_2$ corresponding
to the traversal of the two paths of the interferometer are orthogonal, but
$P_{wave}$ is manifestedly not orthogonal to either one of them. Note that the Hilbert space $H_{wave}$
is not contained within either $H_1$ or $H_2$ since neither 
of the two paths alone is sufficient to cause interference (both
paths are needed). Also note that $\psi_1$ and $\psi_2$ (while 
eigenfunctions of $P_1$ and $P_2$ respectively), are not eigenfunctions
of $P_{wave}$. Hence, unlike in the experiments in ref. \cite{ghose,
rangwala}, $P_{wave}$ does not commute with $P_1$ and $P_2$ and therefore
the wave and particle properties are indeed complementary. Thus,
their simultaneous manifestation will definitely give lie to the
Complementarity Principle.

\medskip

\noindent {\bf {\it Welcher weg} experiments that claim to have validated 
Complementarity or Duality}

The duality principle embodied in Englert's inequality [Equation (1)] was verified in atom interference 
experiments \cite{durr} and perhaps even in recent  experiments conducted with electrons
traversing an Aharonov-Bohm (A-B) quantum interferometer whose one arm contained a quantum dot (QD) with a nearby quantum point contact (QPC) \cite{heiblum1,
heiblum2}. The QD has a non-critical, peripheral role; it merely serves to 
trap an electron traveling that path long enough for the QPC to
detect it. The trapping changes the transmission
probability through the QPC (and hence its conductance) thus 
 allowing ``which path'' detection. 

\begin{figure*}
\epsfxsize=5.8in
\epsfysize=6.3in
\centerline{\epsffile{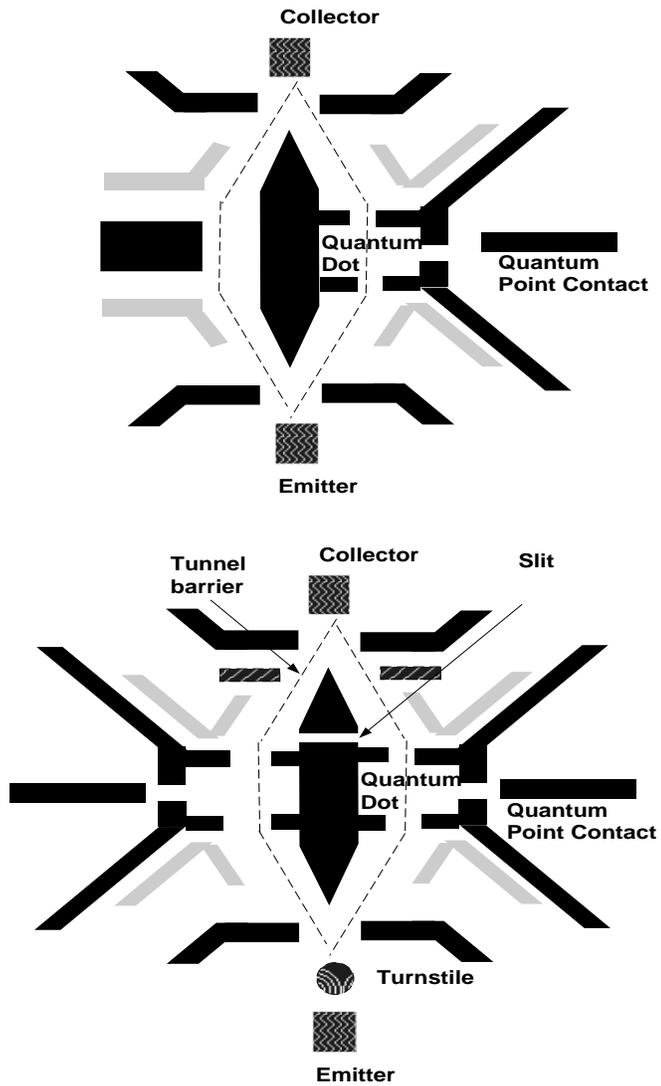}}
\caption{(a) A welcher-weg Aharonov-Bohm interferometry experiment
conducted with electrons to test the Complementarity Principle; (b) a modified welcher-weg 
Aharonov-Bohm interferometry experiment
that can test the Complementarity Principle more rigorously as well as 
test ``quantum erasure''.}
\end{figure*}

The experiment \cite{heiblum1} showed that the A-B interference was diluted if one could even {\it in principle}
detect which path was traversed, irrespective of whether the detection
actually took place.  When the QPC detector
was turned on, the interference peaks were diluted regardless of 
whether one monitored or not any change in the QPC conductance caused
by a fleeting electron in the nearby path. This result, remarkable as it is, does {\it not}  shed any light on the orthodox version of Bohr's
complementarity principle. Validation of the orthodox version 
would require demonstrating {\it complete} vanishing of interference 
along with the demonstration of  perfect anti-coincidence
between {\it two} QPC detectors placed near the two paths of the interferometer. Unfortunately, this was not attempted in the experiment.
Second, the experimental result may be consistent with the Duality Principle [Equation (1)] (like  ref. \cite{durr}), but does 
not quite validate it either (validation is a stronger condition than consistence).
We say this because
turning the QPC on introduces an {\it asymmetry} into the interferometer
(e.g. decrease the transmittivity of one path relative to another) and this alone can cause a dilution of the interference as 
we show in the appendix.

 We suggest 
some modifications to this experiment to test the orthodox version of 
Bohr's complementarity principle. The configuration of the interferometer used in ref. \cite{heiblum1} is shown 
in Fig. 2(a). It was defined on a high mobility two-dimensional 
electron gas using standard split-gate technology. The modifications are
the following:

\begin{enumerate}

\item Introduce electrons one at a time into 
the interferometer
using single charge tunneling  (single electron pump or turnstile) \cite{grabert}. This can be done by 
delineating a small island at the mouth of the emitter and isolating it 
from the emitter with a tunnel barrier whose resistance is much higher 
than $h/e^2$ ($h$ = Planck's constant and $e$ = single electron charge).
Modulating the barrier with an external potential can cause electrons
to be injected one at a time.
There may be other, more complicated, schemes for pumping single electrons
\cite{markus, altshuler} that may be more appropriate depending on 
the measurement approach.

\item Place two tunnel barriers in the 
two arms of the A-B interferometer near the detector $C$. They can also be 
defined by split gates and are shown as striped metal gates in Fig. 2(b). If 
an electron is detected  and it collapses 
to a ``particle'', it {\it cannot} tunnel through the 
barrier and reach the collector C since only waves tunnel and 
particles do not tunnel. Therefore, it must be reflected back into the
emitter. Thereafter it  cannot take the other path to the collector
either because there is a tunnel barrier in that path as well. Consequently
there should be no current. The evolution of the particle nature 
should not only destroy the interference, but actually 
reduce the current to zero. 

\item Place a narrow slit  defined by split gates in between the
QPC detectors and the detector C. This connects the two paths.
Normally this slit is pinched off by the split gates to isolate 
the two paths, but it can be opened (made conducting) by the split gate
voltages, if necessary. This arrangement is relevant to the study of
"quantum erasure" as shown later.
\item Place {\it two} QD-QPC detectors alongside both arms. They serve two
purposes. They can make the interferometer {\it symmetric} and they could also
be
used to demonstrate a perfect anti-coincidence between the 
detection events in the two paths (detection events correspond to a change in the
QPC current when an electron passes by).

\end{enumerate}

With the above modifications, the A-B interferometer can now be 
used to remove all the objections that we raised in relation to
the experiment of ref. \cite{heiblum1}. The interferometer is nominally
symmetric as long as both detectors are turned on. Perfect 
anti-coincidence can be demonstrated if it exists, and the tunnel
barriers in the interferometer arms will conclusively demonstrate 
if the wave nature is present or absent. Thus {\it sharp}
particle- and wave-behavior can be exhibited with no ambiguity.

\medskip

\noindent {\bf Which path determination and quantum erasure}

\medskip

In addition to other differences, a major point of departure between
the experiment of ref. \cite{ghose} and the above experiment is that
the photon experiment purports to demonstrate wave behavior  {\it before} the particle behavior is evidenced in
the transmission path, whereas here the opposite is proposed.
The tunneling gap in Fig. 1(a) {\it precedes} the detector $D_t$. In our experiment (Fig. 2(b)),
the reverse is true; the detector precedes the tunnel barrier.
Thus, simultaneous exhibition of wave- and particle- nature in
our proposed experiment 
would require an electron to first ``become'' a particle
and then be reincarnated as a wave. 
``Which path'' determination does not prohibit this since entanglement
of the electron's wavefunction with that of the detector does not 
cause the entangled pure state to evolve into a mixed state 
irreversibly. We show this 
below.

Let $\psi_1$ and $\psi_2$ be the wavefunctions in the two arms of
the interferometer and $\theta$ is the A-B phase difference 
between the two arms. Then, neglecting multiple 
reflection effects \cite{bandy}, the wavefunction $\psi$ at the 
detector D is
\begin{equation}
\psi = \psi_1 + e^{i \theta} \psi_2
\end{equation}
The current density at the detector is
\begin{equation}
J = {{i e \hbar}\over{2 m^*}} \left [ \psi (\nabla \psi)^* - \psi^* \nabla \psi \right ]
\end{equation}
Thus, for propagating states (plane waves along the direction of 
propagation), Equations (1) and (2) yield
\begin{equation}
J \propto |\psi|^2 =   |\psi_1|^2 +  |\psi_2|^2 + e^{i \theta} \psi_1^* \psi_2
+ e^{-i \theta} \psi_2^* \psi_1
\end{equation}
The last two terms account for the A-B interference.

A fundamental result of quantum measurement theory is that if a detector tries to detect which arm was traversed by the 
interferometer, the wavefunction of the detector becomes 
entangled with that of the electron. The entangled wavefunction  can be
written as 
\begin{equation}
\Phi =  \psi_1 |1> +  e^{i \theta} \psi_2 |2> 
\end{equation}
where the wavefunctions $|1>$ and $|2>$ span the Hilbert space of the 
detector. The state $|1>$ corresponds to detecting the particle in
path 1 and $|2>$ corresponds to detecting the particle in
path 2.

The current density associated with this wave function is
\begin{eqnarray}
J_{entangled} \propto |\Phi|^2 = |\psi_1|^2 <1|1> + |\psi_2|^2 <2|2> \nonumber \\
+ <1|2>e^{i \theta} \psi_1^* \psi_2 + <2|1>e^{-i \theta} \psi_2^* \psi_1
\end{eqnarray}

If the detector is an ``unambiguous'' detector which unambiguously
determines which path is traversed by the particle, then $|1>$ and $|2>$
are orthonormal and hence the interference terms  
vanish in Equation (5) and the wave behavior is lost. This is interpreted as 
dephasing or collapse. 
\begin{equation}
J_{collapsed} \propto  \left [ |\psi_1|^2  + |\psi_2|^2 \right ]
\end{equation}
In truth however, the collapse is
not quite irreversible since if we design an experiment whose 
result is the probability of a particular outcome of the {\it welcher
weg} determination {\it and} finding the detector in the 
symmetric state
$[ |1> + |2>]$, we find
\begin{eqnarray}
J_{erase} & \propto &|   [ <1| + <2|]|\Phi > |^2 \nonumber \\
& = & |  [ <1| + <2|]|[ \psi_1|1> + e^{i \theta} \psi_2|2>] |^2 \nonumber \\
& = & [\psi_1|^2 +  |\psi_2|^2 + e^{i \theta} \psi_1^* \psi_2
+ e^{-i \theta} \psi_2^* \psi_1] \nonumber \\
& = & |\psi_1 + e^{i \theta} \psi_2|^2 = |\Phi|^2
\end{eqnarray}
Thus, the original wavefunction of Equation (4) (along with the interference terms) is regenerated from the entangled wavefunction once we choose to erase the "which path" information.
This is termed ``quantum erasure'' \cite{hillery, scully, kwiat} and 
is possible because the entangled state of Equation (4) is still
a pure state and not a mixed state. Evolution of a pure state into
a mixed state (termed orthodox collapse) \cite{von-neumann} is 
irreversible, but that is not the case here.

We can test the quantum erasure by opening the slit which, in principle, allows conduction between the two paths and hence effectively erases 
the which path information. This can then regenerate the wave nature
and allow tunneling through the tunnel barriers in the two paths.
Consequently, the current (which is ideally zero) with the slit
closed and the QPC detectors on, will rise to a non-zero value when the slit is opened. This 
will be a demonstration of quantum erasure.

In conclusion, we have proposed modifications to some welcher weg
experiments that we believe can rigorously test the Complementarity Principle.

\pagebreak

\noindent {\bf Appendix}

We show that introducing asymmetry between the two paths of an interferometer
degrades the visibility of the interferograms independent of any other effect.

Consider a two-path interferometer. The
current between the source and the detector can be written as (neglecting multiple reflection of the electron
wave between the emitter and collector (due to geometric 
discontinuities) 
\begin{equation}
I = | i_1 + i_2e^{i \theta} |^2 
\end{equation}
where $i_1(i_2)$ is the complex amplitude of the current in path 1(2) 
and we assume that $|i_1|$ = $\alpha |i_2|$ (0 $\leq$ $\alpha \leq$ 1), i.e.
the transmissivity of path 2 is equal to or larger than that of path 1. Thus,
$\alpha$ is a measure of the asymmetry; the farther $\alpha$ is from
unity, the more is the asymmetry. The
angle $\theta$ is the phase difference between the two paths. We now find
\begin{equation}
I = |i|^2[ 1 + \alpha^2 +2 \alpha cos \theta ]
\end{equation}

We can adopt a suitable metric for  the visibility of the oscillation. This
could be the relative modulation $M$ of the current
\begin{equation}
M = {{I_{max} - I_{min}}\over{I_{max}}} = {{4 \alpha}\over{(1 + \alpha)^2}}
\end{equation}
From the above equation, we can see that $M$ is 100\% if $\alpha$ = 1;
otherwise, $M$ decreases as $\alpha$ decreases. Thus, increasing the 
asymmetry in a two-path interferometer degrades the visibility 
of the interference oscillations independent of any other effect.

\bigskip


\end{document}